\documentclass[prl,preprint,showpacs,preprintnumbers,amsmath,amssymb]{revtex4}

\usepackage[dvips]{graphicx}
\usepackage{dcolumn}
\usepackage{bm}
\usepackage{times}
\usepackage{mathptmx}
\usepackage{color}

\begin{document}

\title{Incompatible Magnetic Order in Multiferroic Hexagonal DyMnO$_3$}

\author{C. Wehrenfennig,$^1$ D. Meier,$^1$ Th.\ Lottermoser,$^1$ Th.\ Lonkai,$^2$ J.-U. Hoffmann,$^3$ N. Aliouane,$^3$ D. N. Argyriou,$^3$ and M. Fiebig$^{1,\ast}$}

 \affiliation{$^1$HISKP, Universit\"{a}t Bonn, Nussallee 14-16, 53115 Bonn, Germany}
 \affiliation{$^2$Ganerben-Gymnasium, M\"{u}hlbergstra{\ss}e 65, 74653 K\"{u}nzelsau, Germany}
 \affiliation{$^3$Helmholtz-Zentrum Berlin f\"{u}r Materialien und Energie, Glienicker Stra{\ss}e 100, 14109 Berlin, Germany}

\begin{abstract}
Magnetic order of the manganese and rare-earth lattices according to different symmetry
representations is observed in multiferroic hexagonal \mbox{(h-)} DyMnO$_3$ by optical second
harmonic generation and neutron diffraction. The incompatibility reveals that the $3d$--$4f$
coupling in the h-$R$MnO$_3$ system  ($R$ = Sc, Y, In, Dy -- Lu) is substantially less developed
than commonly expected. As a consequence, magnetoelectric coupling effects in this type of
split-order parameter multiferroic that were previously assigned to a pronounced $3d$--$4f$
coupling have now to be scrutinized with respect to their origin.
\end{abstract}


\date{Dated
\today}
\maketitle

Controlling magnetism by electric fields and (di)electric properties by magnetic fields poses a
great challenge to contemporary condensed-matter physics. Possibly the most fertile source for
such ``magnetoelectric'' cross correlations are compounds with a coexistence of magnetic and
electric long-range order, called multiferroics \cite{Fiebig05a}. In the {\it
joint-order-parameter multiferroics} the magnetic and the ferroelectric order are related to the
same order parameter \cite{Cheong07}. Although this leads to very pronounced magnetoelectric
interactions the improper spontaneous polarization is extremely small ($\ll 1$~$\mu$C/cm$^2$). In
the {\it split-order-parameter multiferroics} magnetic and ferroelectric order emerge
independently. Therefore, they display a technologically feasible proper spontaneous polarization
of $1-100$~$\mu$C/cm$^2$. Aside from the ambient multiferroic BiFeO$_3$ \cite{Palai08} a
split-order-parameter system at the center of intense discussion is hexagonal (h-) $R$MnO$_3$ with
$R$ = Sc, Y, In, Dy--Lu \cite{Yen07}. The system displays a variety of multiferroic phases and a
``giant'' magnetoelectric effect \cite{Lottermoser04}. The availibility of as many as nine
h-$R$MnO$_3$ compounds is ideal for investigating the role played by magnetic $3d$--$4f$
interactions in the manifestation of magnetoelectric effects, a key question of multiferroics
research.

Thus far, it is assumed that the $3d$--$4f$ interaction in the h-$R$MnO$_3$ system is strong with
a rigid correlation between the magnetic Mn$^{3+}$ and $R^{3+}$ order
\cite{Fabreges08,Harikrishnan09,Lonkai02,Sugie02,Munawar06}. The Mn$^{3+}$--$R^{3+}$ exchange
paths are affected by the ferroelectric distortion of the unit cell so that it is expected that,
reminiscent of the orthorhombic manganites \cite{Cheong07}, the $3d$--$4f$ interaction has
substantial impact on the magnetoelectric behavior, including giant magnetoelectric
\cite{Lottermoser04} and magnetoelastic \cite{Lee08,Fabreges09} effects.

In this Letter we show that the $3d$--$4f$ coupling in h-$R$MnO$_3$ is substantially less
developed than assumed up to now. This is concluded from optical second harmonic generation (SHG)
and neutron-diffraction data revealing that the Mn$^{3+}$ spins and the Dy$^{3+}$ spins in
h-DyMnO$_3$ order according to different symmetry representations unless magnetization fields are
present. This has extensive consequences for the magnetic structure and the magnetoelectric
interactions in the multiferroic h-$R$MnO$_3$ system that are discussed in detail.

The h-$R$MnO$_3$ compounds display ferroelectric ordering at $T_{\rm C}=650-990$~K,
antiferromagnetic Mn$^{3+}$ ordering at $T_{\rm N}=66-130$~K \cite{Yen07}, and, for $R$ = Dy--Yb,
magnetic $R$$^{3+}$ ordering and reordering at $T_{\rm N}$ and $4-8$~K, respectively
\cite{Yen07,Fabreges08,Lonkai02,Nandi08a,Sugie02}. The spontaneous polarization is
5.6~$\mu$C/cm$^2$ and directed along $z$. Frustration leads to a variety of triangular
antiferromagnetic structures of the Mn$^{3+}$ spins in the basal $xy$ plane. In contrast, the
$R^{3+}$ sublattices order Ising-like along the hexagonal $z$ axis. The possible magnetic
structures of the Mn$^{3+}$ and Dy$^{3+}$ lattices correspond to four one-dimensional
representations \cite{footnote2} that are compared in Table~\ref{tab1}. An extensive investigation
of the magnetic $R^{3+}$ order commenced only recently
\cite{Fabreges08,Lonkai02,Nandi08a,Nandi08b} and revealed that in h-HoMnO$_3$ and h-YbMnO$_3$ the
Mn$^{3+}$ and the $R^{3+}$ ordering occurs according to the same magnetic representation.

The compound with the smallest $R^{3+}$ ion in the h-$R$MnO$_3$ series is h-DyMnO$_3$ which was
grown for the first time only recently \cite{Ivanov06}. The magnetic structure of the Dy$^{3+}$
lattice was investigated by resonant x-ray diffraction and magnetization measurements
\cite{Nandi08a,Ivanov06}. Its magnetic point group was found to be
$P\underline{6}_3c\underline{m}$ in the interval $10~{\rm K}<T<T_{\rm N}$ (``high-temperature
range''). At $T<10$~K (``low-temperature range'') or above a critical magnetic field applied along
$z$ it changes to $P6_3\underline{cm}$. In this work we focus on the determination of the
complementary Mn$^{3+}$ order, but for confirmation the Dy$^{3+}$ order is also verified.

SHG is described by the equation $P_i(2\omega) = \epsilon_0 \chi_{ijk} E_j(\omega) E_k(\omega)$.
An electromagnetic light field $\vec{E}$ at frequency $\omega$ is incident on a crystal, inducing
a dipole oscillation $\vec{P}(2\omega)$, which acts as source of a frequency-doubled light wave of
the intensity $I_{\rm SHG}\propto|\vec{P}(2\omega)|^2$. The susceptibility $\chi_{ijk}$ couples
incident light fields with polarizations $j$ and $k$ to a SHG contribution with polarization $i$.
The magnetic and crystallographic symmetry of a compound uniquely determines the set of nonzero
components $\chi_{ijk}$ \cite{Birss66,Fiebig05b}. In turn, observation of $\chi_{ijk}\neq 0$
allows one to derive the the magnetic symmetry and structure.

The h-DyMnO$_3$ samples were obtained by the floating zone technique and verified for the absence
of twinning and secondary phases by Laue diffraction. SHG reflection spectroscopy with 120-fs
laser pulses was conducted on polished $z$-oriented crystals in a $^4$He-operated cryostat
generating magnetic fields of up to 8~T \cite{Fiebig05b}. Neutron diffraction in the $(h0l)$ plane
was conducted at the E2 beamline of the Helmholtz-Zentrum at a wavelength of {2.39~\AA} with the
sample mounted in a $^3$He/$^4$He dilution insert.

In Table~\ref{tab1} the calculated selection rules identifying the magnetic structure of the
Mn$^{3+}$ and Dy$^{3+}$ lattice are listed. SHG is only sensitive to the Mn$^{3+}$ order with
$\chi_{xxx}$ and $\chi_{yyy}$ as independent tensor components \cite{Birss66}. Neutron diffraction
can probe the magnetic moments of Mn$^{3+}$ and Dy$^{3+}$. The (100) and (101) reflections lead to
very clear selection rules so that we restrict the discussion to them. The contributions by the
Mn$^{3+}$ and Dy$^{3+}$ lattice to these reflections were separated by setting the magnetic moment
of the other lattice, respectively, to zero in the computations done with {\it Simref~2.6}
\cite{Lonkai02}.

\begin{figure}[h]
\begin{center}
\includegraphics[width=8.7cm,keepaspectratio,clip]{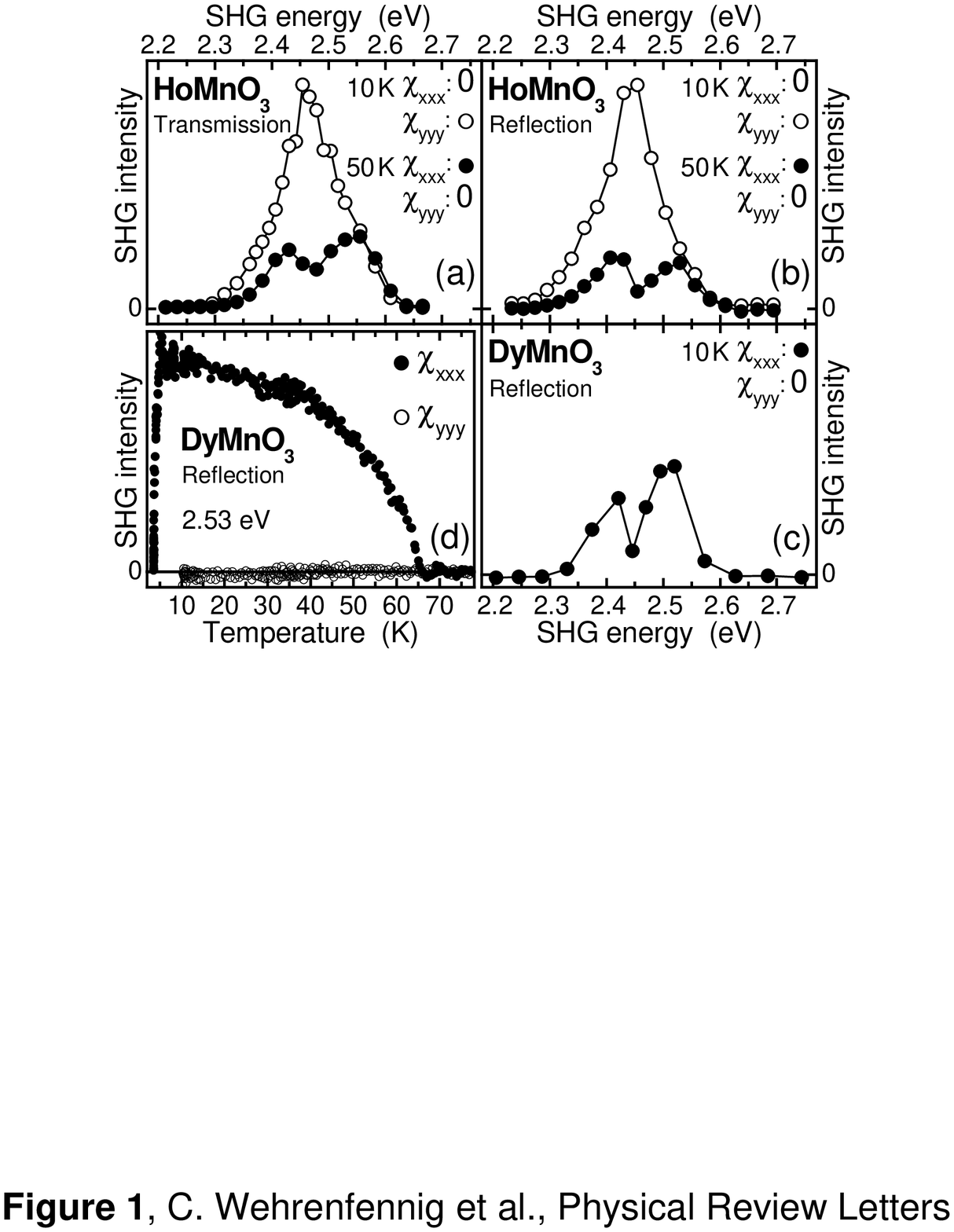}
\end{center}
\caption{\label{fig1} Spectral, polarization, and temperature dependence of SHG in h-$R$MnO$_3$
compounds. A ``0'' indicates zero SHG intensity. (a, b) SHG spectra of h-HoMnO$_3$ measured in (a)
transmission and (b) reflection with a fs laser. (c) SHG spectra of h-DyMnO$_3$ measured in
reflection with a fs laser. (d) Temperature dependence of the signal in (c).}
\end{figure}

We first focus on the high-temperature range and measurements at zero magnetic field.
Figure~\ref{fig1} shows the analysis of the magnetic structure of the Mn$^{3+}$ lattice by SHG
spectroscopy. Because of the large optical absorption the SHG data on h-DyMnO$_3$ cannot be taken
with the standard transmission setup and ns laser pulses \cite{Fiebig05b} --- in contrast to all
other h-$R$MnO$_3$ compounds. Instead the reflected SHG signal was measured with a fs laser
system. With fs laser pulses, higher-order SHG contributions, incoherent multiphoton processes,
and ultrafast nonequilibrium effects can easily obscure any magnetically induced SHG
\cite{Fiebig05b,Hillebrands08}. In the first step we therefore had to verify to what extent SHG is
still a feasible probe for the magnetic structure. We chose h-HoMnO$_3$ for this test since it
allows us to compare transmission and reflection data. Figure~\ref{fig1}(a) shows the SHG
transmission spectrum taken at two different temperatures with a fs laser system. Aside from a
minor decrease of resolution the fs laser pulses lead to the same SHG spectra as the ns laser
pulses \cite{Fiebig05b}. With SHG from $\chi_{xxx}$ at 50~K and from $\chi_{yyy}$ at 10~K we
identify the $P\underline{6}_3\underline{c}m$ and the $P\underline{6}_3c\underline{m}$ structure,
respectively, on the basis of Table~\ref{tab1}. Note that the spectral dependence is also
characteristic for the respective phases \cite{Iizuka01}. Figure~\ref{fig1}(b) shows the
corresponding SHG reflection spectra. Aside from a 98\% decrease of the SHG yield the result
remains unchanged. Hence, the fs reflection data are well suited for identifying the magnetic
phase of h-$R$MnO$_3$ by SHG.

Figures~\ref{fig1}(c) and \ref{fig1}(d) show the spectral and temperature dependence of the SHG
signal in h-DyMnO$_3$ in a fs reflection experiment. Comparison with Fig.~\ref{fig1}(b) and
Table~\ref{tab1} clearly reveals $P\underline{6}_3\underline{c}m$ as magnetic symmetry of the
Mn$^{3+}$ lattice in the high-temperature range up to $T_{\rm N}=66$~K. This is an utterly
surprising result because $P\underline{6}_3\underline{c}m$ does not match the
$P\underline{6}_3c\underline{m}$ symmetry of the Dy$^{3+}$ lattice proposed in
Ref.~\onlinecite{Nandi08a}. We therefore sought additional confirmation by neutron diffraction.

\begin{figure}[h]
\begin{center}
\includegraphics[width=8.7cm,keepaspectratio,clip]{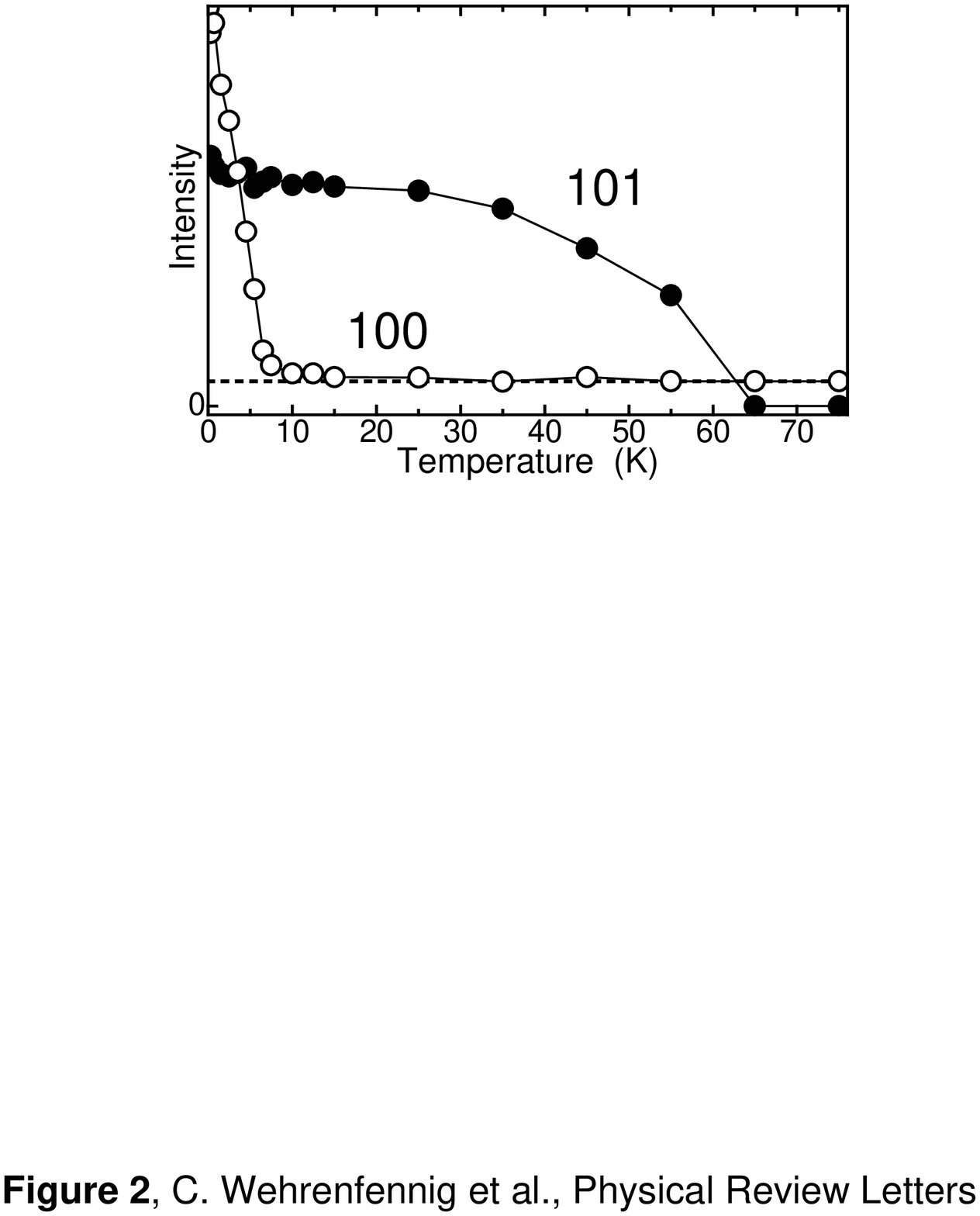}
\end{center}
\caption{\label{fig2} Temperature dependence of the (101) and (100) reflections of h-DyMnO$_3$
measured in a cooling run on the sample from Fig.~\protect\ref{fig1}. The dashed line marks the
offset due to crystallographic contributions.}
\end{figure}

In Fig.~\ref{fig2} we show the temperature dependence of the (101) and (100) reflections of
h-DyMnO$_3$. Note that while the (101) reflection is magnetically induced the (100) reflection in
the high-temperature range is entirely due to crystallographic contributions
--- its magnetic intensity is zero. According to Table~\ref{tab1}, this is only possible if
the magnetic symmetry of the Mn$^{3+}$ lattice is either $P\underline{6}_3\underline{c}m$ or
$P6_3\underline{cm}$ of which the latter corresponds to a ferromagnetic state which is ruled out
by magnetization measurements \cite{Ivanov06,Harikrishnan09,Nandi08a}. Table~\ref{tab1} further
shows that the magnetic symmetry of the Dy$^{3+}$ lattice can only be
$P\underline{6}_3c\underline{m}$ which confirms the magnetic structure proposed earlier
\cite{Nandi08a}.

We thus conclude that three independent experimental parameters, i.e., SHG polarization, SHG
spectrum, and neutron diffraction intensity, confirm $P\underline{6}_3\underline{c}m$ as magnetic
symmetry of the Mn$^{3+}$ lattice in the high-temperature range of h-DyMnO$_3$. This is in
striking contrast to the known \cite{Nandi08a} $P\underline{6}_3c\underline{m}$ symmetry of the
Dy$^{3+}$ lattice which is confirmed by our neutron data. Hence, the Mn$^{3+}$ and the Dy$^{3+}$
order are ``incompatible'' in the sense of belonging to different magnetic representations.

Although it is not unusual that magnetic order is parametrized by more than one representation it
is most remarkable that this occurs in the h-$R$MnO$_3$ system. Up to now, research on this system
was based on the assumption that a pronounced coupling between the Mn$^{3+}$ and $R^{3+}$ lattices
is a central mechanism in determining its magnetoelectric and multiferroic properties
\cite{Fabreges08,Harikrishnan09,Lonkai02,Sugie02,Munawar06,Lottermoser04,Lee08}. Accordingly,
ordering of the $3d$ and $4f$ lattices according to a single representation was implied to be
compulsory. However, the incompatibility observed here shows that the $3d$--$4f$ coupling must be
distinctly less developed than commonly assumed: It competes with other effects of comparable
magnitude that can be associated to a different magnetic structure.

Consequently, a variety of phenomena that were related to a pronounced $3d$--$4f$ coupling have to
be scrutinized. This includes the $P\underline{6}_3\underline{c}m\to
P\underline{6}_3c\underline{m}$ reorientation of the Mn$^{3+}$ lattice in HoMnO$_3$
\cite{Fiebig03,Yen07}, which plays a role in the emergence of a giant magnetoelectric effect
\cite{Lottermoser04}. Supplementary (or alternative) to Mn$^{3+}$--Ho$^{3+}$ exchange the
temperature dependence of the magnetic anisotropy or magnetoelastic effects \cite{Fabreges09} may
be responsible for the reorientation. As another issue, $R^{3+}$ ordering in the high-temperature
range cannot be due to straightforward induction by the Mn$^{3+}$ order since this would
inevitably lead to compatible Mn$^{3+}$ and $R^{3+}$ order. Finally, we cannot confirm that the
Mn$^{3+}$ order is directly determined by the size of the $R^{3+}$ ion and, thus, by the
$R^{3+}$--O$^{2-}$--Mn$^{3+}$ bond angle \cite{Kozlenko07}. If this were the case, the Mn$^{3+}$
spins in DyMnO$_3$ with the smallest $R^{3+}$ ion of the h-$R$MnO$_3$ system would order according
to the $P\underline{6}_3c\underline{m}$ structure already observed in HoMnO$_3$ and YMnO$_3$.
Likewise, it can be excluded that the unusual magnetic phase diagram and magnetoelectric
properties established for HoMnO$_3$ \cite{Fiebig03,Yen07,Vajk05} continue towards rare-earth
h-$R$MnO$_3$ compounds with a smaller $R^{3+}$ radius than Ho$^{3+}$. These observations, too,
corroborate the relative independence of the Mn$^{3+}$ and $R^{3+}$ lattices.

In the low-temperature range and in a magnetic field applied along $z$ the magnetic order of the
Dy$^{3+}$ spins changes from $P\underline{6}_3c\underline{m}$ to $P6_3\underline{cm}$
\cite{Nandi08a,Ivanov06}. In the following the effect of this transition on the Mn$^{3+}$ spins is
investigated. The emerging intensity of the (100) peak in Fig.~\ref{fig2} reflects the transition
of the Dy$^{3+}$ lattice to the $P6_3\underline{cm}$ phase in agreement with Table~\ref{tab1}.
However, it also obscures the Mn$^{3+}$-related contributions so that we revert to SHG
measurements for a unique identification of the Mn$^{3+}$ order.

\begin{figure}[h]
\begin{center}
\includegraphics[width=8.7cm,keepaspectratio,clip]{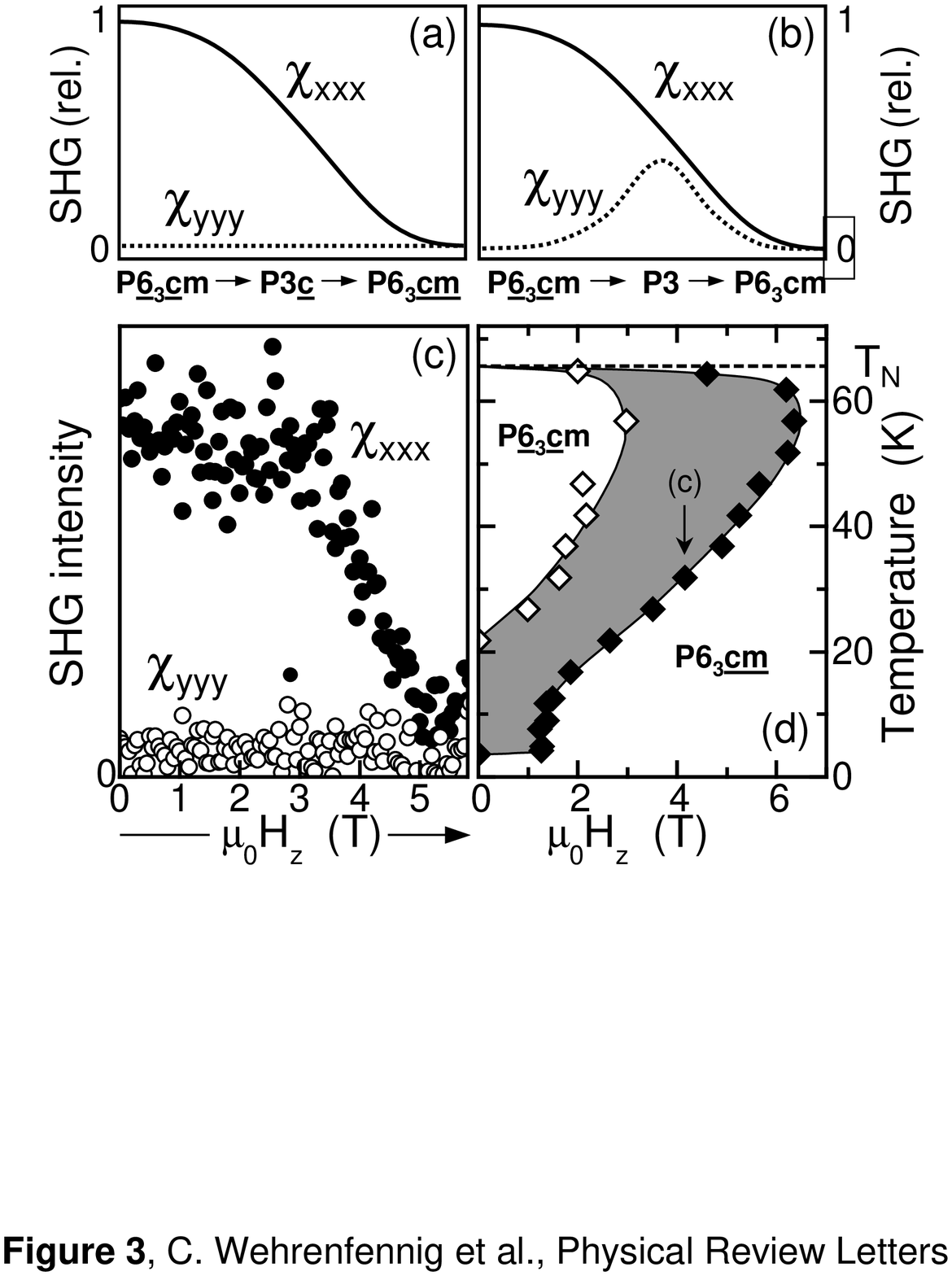}
\end{center}
\caption{\label{fig3} Phase diagram of the magnetic Mn$^{3+}$ order in h-DyMnO$_3$ exposed to a
magnetic field along $z$. (a, b) Sketch of the SHG tensor contributions in the course of the  spin
reorientation towards the (a) $P6_3\underline{cm}$ and (b) $P6_3cm$ phase. (c) Exemplary field
dependence of the SHG signal at 32~K measured in a field-increasing run. (d) Phase diagram derived
from measurements as in (c). Temperature-increasing and -decreasing runs yield the boundaries on
the right- and left-hand side of the gray area, respectively. The boundary values correspond to
observation of 50\% of the SHG yield of the $P\underline{6}_3\underline{c}m$ phase.}
\end{figure}

Figure~\ref{fig1}(d) and Fig.~\ref{fig3}(c) show that the SHG intensity is quenched in the
low-temperature range and in a magnetic field. According to Table~\ref{tab1} this indicates a
transition to either the $P6_3\underline{cm}$ or the $P6_3cm$ phase. According to
Figs.~\ref{fig3}(a) and \ref{fig3}(b) the $P\underline{6}_3c\underline{m}\to P6_3\underline{cm}$
transition passes through a state with $P3\underline{c}$ symmetry for which $\chi_{xxx}\neq 0$,
$\chi_{yyy}=0$ while the $P\underline{6}_3c\underline{m}\to P6_3cm$ transition passes through a
state with $P3$ symmetry for which $\chi_{xxx}\neq 0$, $\chi_{yyy}\neq 0$ \cite{Birss66}.
Figure~\ref{fig3}(c) reveals that the former scenario is realized. This is confirmed by the phase
diagram in Fig.~\ref{fig3}(d). Data points denote the magnetic field at which a magnetic phase
transitions are observed with the gray area marking the difference between field-increasing and
-decreasing runs. Qualitatively, the same phase diagram was obtained on h-$R$MnO$_3$ with $R$ =
Er, Tm, Yb, all of which exhibit a $P\underline{6}_3c\underline{m}\to P6_3\underline{cm}$
transition of the Mn$^{3+}$ lattice in a magnetic field whereas the phase diagram of h-HoMnO$_3$,
in which this transition does not occur, is different. We conclude that in the low-temperature
range and in a magnetic field along $z$ the magnetic symmetry of the Mn$^{3+}$ lattice in
h-DyMnO$_3$ is $P6_3\underline{cm}$ and, thus, compatible to the magnetic symmetry of the
Dy$^{3+}$ lattice.

Nevertheless this compatibility is not a compulsory indication for an enhanced $3d$--$4f$ exchange
in the low-temperature range. $P6_3\underline{cm}$ is a ferromagnetic point group and the
observation that a magnetic field supports the transition into this state indicates that the
magnetic field energy rather than Mn$^{3+}$--Dy$^{3+}$ exchange coupling may be responsible for
the $P\underline{6}_3\underline{c}m\to P6_3\underline{cm}$ transition of the Mn$^{3+}$ lattice.
The field is generated internally by the ferromagnetic order of the Dy$^{3+}$ lattice and
supported externally by the applied magnetic field. Accordingly, the gray area in
Fig.~\ref{fig3}(d) does not show the hysteresis of a first-order transition of the Mn$^{3+}$
lattice but rather the response of the Mn$^{3+}$ lattice to the magnetizing field exerted by the
Dy$^{3+}$ lattice when it is ferromagnetically ordered.

We thus observed that the magnetic order of the manganese and the rare-earth sublattices in the
h-$R$MnO$_3$ system can be incompatible from the point of view of symmetry. In h-DyMnO$_3$
Mn$^{3+}$ ordering according to the $P\underline{6}_3\underline{c}m$ symmetry and Dy$^{3+}$
ordering according to the $P\underline{6}_3c\underline{m}$ symmetry is revealed. The
incompatibility demonstrates that the $3d$--$4f$ interaction in the h-$R$MnO$_3$ series is
distinctly less developed than assumed up to now. Even when the incompatibility is overcome by an
internal or external magnetization field, pronounced Mn$^{3+}$--Dy$^{3+}$ exchange does not have
to be involved.

As a consequence, a variety of magnetoelectric coupling effects in the h-RMnO$_3$ system that were
previously assigned to pronounced $3d$--$4f$ coupling have now to be scrutinized with respect to
their origin. Apparently, interactions {\it within} the Mn$^{3+}$ or $R^{3+}$ lattices as well as
separate crystallography, anisotropy, and frustration effects play a greater role than expected up
to now. In particular, externally induced polarization \cite{Lottermoser04} or magnetization
\cite{Fiebig03} fields rather than the Mn$^{3+}$--$R^{3+}$ exchange may contribute substantially
to the magnetoelectric response.

C.W., D.M., Th.L., and M.F. thank the DFG (SFB 608) for subsidy. D.N.A. thanks the DFG for
financial support under Contract No. AR 613/1-1.


\newpage


\begin{table}
\caption{\label{tab1} Selection rules for SHG and neutron diffraction distinguishing between the
four one-dimensional symmetry representations of the crystallographic space group $P6_3cm$ of
h-$R$MnO$_3$ \protect\cite{footnote2}.}

\vspace{1.5cm}

\begin{tabular}[t]{l|c||r|r|r|r}
 \multicolumn{2}{l||}{Representation} & $\Gamma_1$, A$_1$ & $\Gamma_2$, A$_2$ & $\Gamma_3$, B$_1$ & $\Gamma_4$, B$_2$\\
 \multicolumn{2}{l||}{Symmetry} & $P6_3cm$ & $P6_3\underline{cm}$ & $P\underline{6}_3c\underline{m}$ & $P\underline{6}_3\underline{c}m$\\
 \hline
 SHG & $\chi_{xxx}$ & 0 & 0 & 0 & $\neq 0$\\
 (Mn$^{3+}$) & $\chi_{yyy}$ & 0 & 0 & $\neq 0$ & 0\\
 \hline
 Neutron\ & (100) & 58\% & 0 & 62\% & 0\\
 (Mn$^{3+}$) & (101) & 17\% & 100\% & 17\% & 93\%\\
 \hline
 Neutron\ & (100) & 0 & 100\% & 0 & 75\%\\
 (Dy$^{3+}$) & (101) & 100\% & 0 & 2\% & 0\\
\end{tabular}

\vspace{1.5cm}

\end{table}

Symmetries refer to the magnetic structure of the Mn$^{3+}$ or Dy$^{3+}$ {\it sublattice}; the
overall symmetry is determined by the intersection of the sublattice symmetries. Sketches of the
magnetic structures of the Mn$^{3+}$ and Dy$^{3+}$ sublattices are given in
Refs.~\protect\onlinecite{Lottermoser04} and \protect\onlinecite{Nandi08a}. SHG selection rules
were derived from Ref.~\protect\onlinecite{Birss66}. The neutron diffraction yield contributed by
the Mn$^{3+}$ and the Dy$^{3+}$ order, respectively, was calculated as described in the text. The
largest value obtained for each sublattice corresponds to 100\%.
\newpage


\begin{thebibliography}{99}

\bibitem[*]{dummy} Email address: fiebig@hiskp.uni-bonn.de

\bibitem{Fiebig05a}
M. Fiebig,
J.\ Phys.\ D {\bf 38}, 123 (2005).

\bibitem{Cheong07}
S.-W. Cheong and M. Mostovoy,
Nat.\ Mater.\ {\bf 6}, 13 (2007).

\bibitem{Palai08}
R. Palai {\it et al.}, 
Phys.\ Rev.\ B {\bf 77}, 014110 (2008).

\bibitem{Yen07}
F. Yen {\it et al.}, 
J.\ Mater.\ Res.\ {\bf 22}, 2163 (2007).

\bibitem{Lottermoser04}
Th.\ Lottermoser {\it et al.}, 
Nature {\bf 430}, 541 (2004).

\bibitem{Fabreges08}
X. Fabr\`{e}ges {\it et al.}, 
Phys.\ Rev.\ B {\bf 78}, 214422 (2008).

\bibitem{Harikrishnan09}
S. Harikrishnan {\it et al.}, 
J.\ Phys.: Condens.\ Matter {\bf 21}, 096002 (2009).

\bibitem{Lonkai02}
Th. Lonkai, D. Hohlwein, J. Ihringer, and W. Prandl,
Appl.\ Phys.\ A, {\bf 74}, S843 (2002).

\bibitem{Sugie02}
H. Sugie, N. Iwata, and K. Kohn,
J.\ Phys.\ Soc.\ Jpn.\ {\bf 71}, 1558 (2002).

\bibitem{Munawar06}
I. Munawar and S. H. Curnoe,
J.\ Phys.: Condens.\ Matter {\bf 18}, 9575 (2006).

\bibitem{Lee08}
S. Lee {\it et al.},
Nature {\bf 451}, 805 (2008).

\bibitem{Fabreges09}
X. Fabr\`{e}ges {\it et al.}, 
Phys.\ Rev.\ Lett.\ {\bf 103}, 067204 (2009).

\bibitem{Nandi08a}
S. Nandi {\it et al.},
Phys.\ Rev.\ B {\bf 78}, 075118 (2008).

\bibitem{footnote2}
The four one-dimensional representations consistently explane our data so that two more
two-dimensional representations denoting lower symmetries
\protect\cite{Nandi08a,Fiebig03,Fabreges08} are omitted.

\bibitem{Nandi08b}
S. Nandi {\it et al.}, 
Phys.\ Rev.\ Lett. {\bf 100}, 217201 (2008).

\bibitem{Ivanov06}
V. Yu. Ivanov {\it et al.}, 
Phys.\ Solid State {\bf 48}, 1726 (2006).

\bibitem{Birss66}
R. R. Birss, {\it Symmetry and Magnetism}, (North Holland, Amsterdam, 1966).

\bibitem{Fiebig05b}
M. Fiebig, V. V. Pavlov, and R. V. Pisarev,
J.\ Opt.\ Soc.\ Am.\ B {\bf 22}, 96 (2005).

\bibitem{Hillebrands08}
B. Hillebrands,
J.\ Phys.\ D: Appl.\ Phys.\ {\bf 41}, 160301 (2008).

\bibitem{Iizuka01}
T. Iizuka-Sakano, E. Hanamura, and Y. Tanabe,
J.\ Phys.: Condens.\ Matter {\bf 13}, 3031 (2001).

\bibitem{Fiebig03}
M. Fiebig, Th.\ Lottermoser, and R. V. Pisarev,
J.\ Appl.\ Phys. {\bf 93}, 8194 (2003).

\bibitem{Kozlenko07}
D. P. Kozlenko {\it et al.}, 
J.\ Phys.: Condens.\ Matter {\bf 19}, 156228 (2007).

\bibitem{Vajk05}
O. P. Vajk {\it et al.}, 
Phys.\ Rev.\ Lett.\ {\bf 94}, 087601 (2005).

\end{thebibliography}
\end{document}